# Electrical switching and memory phenomena observed in redox-gradient dendrimer sandwich devices

**JianChang Li, Silas C. Blackstock, and Greg J. Szulczewski**

The Center for Materials for Information Technology and Department of Chemistry, The University of Alabama, Tuscaloosa, AL 35487-0209

We report on the fabrication of dendrimer sandwich devices with electrical switching and memory properties. The storage media is consisted of a redox-gradient N,N'''-(1,4-phenylene)bis(N,N'N''N'',N''-pentakis(4-methoxyphenyl)-1,3,5-benzenetriamine) (4AAPD) dendrimer layer sandwiched between two N,N'-diphenyl-N,N'-bis(3-methylphenyl)-1,1'-biphenyl-4,4'-diamine (TPD) thin films. The TPD thin films are used as potential barriers. The 4AAPD layer acts as potential well where redox-state changes and consequent electrical transitions of the embedded 4AAPD dendrimer molecules are expected to be effectively triggered and retained, respectively. Experimental results indicated that electrical switching could be reproducibly obtained in such dendrimer sandwiches upon a threshold bias voltage. After switching, the device conductivity could be increased more than three orders of magnitude, which can keep stable for several days in ambient conditions. Our work demonstrates the possibility of using solid-state redox-gradient dendrimer films as hopeful information storage media.

Recently, switchable organic materials have drawn much attention because of their potential application to electronic and information storage devices with high density, low-cost, and flexible features. External pulse voltage is a quit simple and effective stimulus to induce switching in many information storage systems. So far, electrical switching has been observed in polymers,[1] LB films,[2] and small π-conjugated organic molecules.[3,4] Dendrimers, as highly branched macromolecules constructed with an interior core and successive radially branched shell, have attracted intense interest from the viewpoint of potential application to organic electronics, optical devices, and drug delivery.[5,6,7] Some articles have been reported on the synthesis and characterizations of optical, electrochemical, and electro-optical switchable dendrimers.[8,9,10,11] However, these reports are all based on the results detected from dendrimer solutions or films dipped in electrolytes. Due to the influence of the solvent and electrolytes, many undesirables still exist. For instance, they might be difficult to arrange in arrays and to integrate with present silicon-based electronics, which will in turn limit their potential applications. To circumvent such problems, it would be desirable to develop some unique dendrimer molecules for bistable dendrimer-based solid-state devices, which have been inadequately reported to date.

In this letter, we report on the electrical switching of a kind of redox-gradient dendrimer sandwich devices. We demonstrate how a dendrimer thin solid layer could be used as electrical switches. Figure 1 shows the schematic molecular structures of the materials used. The 4AAPD (N,N'''-(1,4-phenylene)bis(N,N'N'N'',N''-pentakis(4-methoxyphenyl)-1,3,5-benzenetriamine)) dendrimer was designed to have an easy-oxidized "core" and harder oxidized "shell" structure. The molecular synthesis, film growth, and electrochemical characterizations have been reported elsewhere.[12,13,14] The unique redox-gradient and core-shell structure may endow 4AAPD molecules suitable for information storage media and other promising applications. The device is typically constructed by inserting a 4AAPD thin layer between two TPD (N,N'-diphenyl-N,N'-bis(3-methylphenyl)-1,1'-biphenyl-4,4'-diamine) thin films. TPD was used as purchased from Aldrich. The device design is based on the redox switchable property of the embedded 4AAPD dendrimer molecules and the potential barrier character of the TPD films. We expect such device structure could both easily induce and greatly enhance the redox-state transitions of the embedded dendrimer molecules so as to obtain stable electrical switching and memory properties.

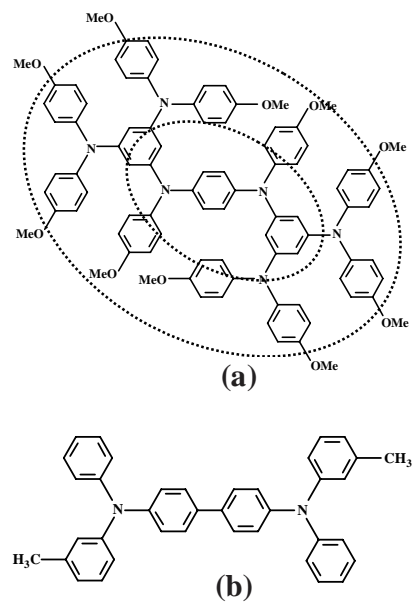

Fig. 1 Schematic chemical structure of (a) 4AAPD and (b) TPD. The core-shell structure of the 4AAPD dendrimers is sketched out by the dashed ovals.

Crossbar device structure was used in this study. First, a 100-nm-thick Ag film was vapor deposited onto clean glass substrate through a contact mask at the pressure of $6\times10^{-7}$ Torr, which defines four 1-mm-wide parallel electrodes. Then remove the mask, the bottom TPD, the middle 4AAPD, and the top TPD layers were consequently evaporated onto the substrates. Finally, load the shadow mask for the top electrodes that is perpendicular to that of the bottom electrodes. The width and thickness of the top electrodes is 1 mm and 40 nm, respectively. So the size our devices is $1\times1\text{mm}^2$. The dc current-voltage (I-V) characteristics of the devices were measured by using a PAR-273 potentiostat (EG&G Instruments) with a 10 KΩ protecting resistor in series. Without special indication, the two TPD layers were kept at the same thickness, the electrodes were used the same metal.

Electrical switching was reproducibly observed in 4AAPD and TPD sandwich devices. Figure 2(a) shows the I-V curves of a typical 4AAPD and TPD sandwich device just before and after switching (device parameters given in the inset). It is shown that the device initially exhibits a high-resistance state in the low bias voltage region. When the applied voltage surpasses a threshold of about 0.5 V, the device was suddenly switched to a low-resistance state. The device conductivity was thus increased about three orders of magnitude. Applying reverse bias voltage cannot "erase" the switched device back to its high-resistance state. A number of devices were examined, the electrical switching is found to be independent of the direction of the applied voltage. The switch threshold-applied voltage ($V_T$) varied from 0.4 to 0.8 V for different devices. Any bias voltage,

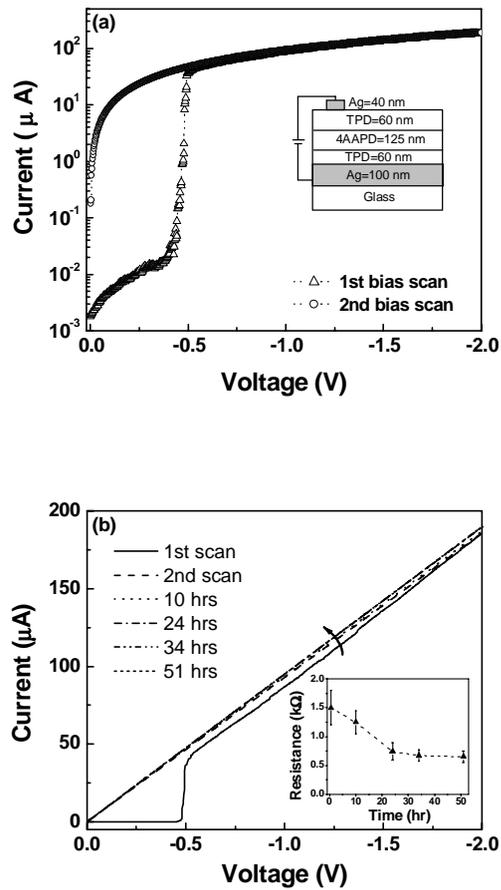

**Fig. 2 (a) Typical I-V characteristics of a TPD and 4AAPD sandwich device just before and after switching. (b) I-V curves recorded at different times for the switched device. Inset shows the resistance changes averaged from a number of switched devices.**

lower than 0.3 V, will not induce the electrical switching. A bias voltage, higher than 5 V, will lead most of the un-switched devices to short (it is noteworthy that those devices, pre-switched by $V_T$, were usually stable under such higher bias voltage for several cycles).

The stability of the switched devices was studied by keeping track of the resistance changes of a number of switched devices for several days. As shown in Fig. 2(b), it was indicated that the low-resistance state of the switched devices could retained for more than 50 hours in ambient conditions (some devices even still kept their low-resistance state after one week). Usually, there exists a resistance decreasing in the first few hours (see inset of Fig. 2(b)), which may be resulted from some exterior influences. Organic films usually tend to absorb water from the air, which would make the film resistance decrease.[15] In our case, when the absorption reaches saturation, the device resistance kept stable at 0.5 ~ 0.8 kΩ. Interestingly, some switched devices could recover its original high-impendence state after being kept in the ambient conditions for a few hours. Under a $V_T$, such devices could switch to low-resistance state again, proposing that some reversible transitions may have taken place in these devices.

The electrical switching effect was found to be strongly dependent on the device structure. Similar electrical switching behavior could be easily observed in either Ag/TPD/4AAPD/TPD/Ag sandwich or Ag/TPD/4AAPD/TPD/4AAPD/TPD/Ag multilayer devices upon a $V_T$. Fig. 3 (a) shows the typical I-V curves of a multilayer device just before and after switching. In the first bias scan, the device initially shows a high resistant state, then abruptly switches to a low resistant state when the applied voltage reaches about 1.8 V. After switching, the device conductivity is increased about two orders of magnitude, which was stable even under different pulse voltages such as (0.8 V, 60 s), (1.5V, 60 s), and (2.5 V, 120 s), respectively (see inset of Fig. 3(a)). The average $V_T$ for the 50 nm multilayer devices is about 0.9 V, which is consistent with that of the sandwiches. As indicated by the arrows, multi-step switching process was interestingly noted in these multilayer devices. Similar behavior was repeatedly observed in most of the multilayer devices. This phenomenon is dependent on the multilayer structure, because without it, no such multi-step switching could be observed. Fig.3 (b) shows the typical I-t curve of a multilayer device before and after switching by applying a -2.5V, 120s pulse voltage. It was shown that the switching time is shorter than 0.5 s (due to our instrument limitation, actuate switching time is still not clear).

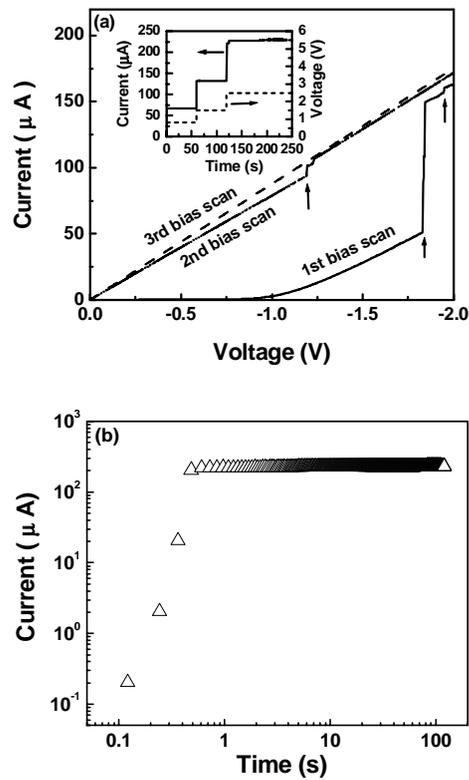

**Fig. 3 (a) Typical I-V characteristics of a Ag/TPD/4AAPD/TPD/4AAPD/TPD/Ag (each layer is 10-nm-thick) multilayer device just before and after switching. Inset demonstrates that the low-resistance state kept stable under different pulse voltages. (b) Representative I-t curves of a typical TPD and 4AAPD multilayer device just before and after switching by applying a -2.5 V, 120s pulse voltage.**

We intentionally fabricated a series of sandwich and multilayer devices with thickness of each TPD and 4AAPD layer from 10 to 80 nm and 10 to 120 nm, respectively. Statistically, ~ 25 % of the 600 devices, we measured, exhibit similar switching effect under a threshold-applied voltage. All the others are shorted, which may be due to device local microstructure defects. However, no resemble switching phenomena could be observed in those devices with either a neat layer of TPD or 4AAPD. Moreover, no switching behavior could be detected in Ag/TPD/4AAPD/Ag bilayer devices fabricated at the same conditions. Totally, about 900 single and bilayer devices were tested. Whatever the TPD and 4AAPD single or bilayer thickness is, these devices usually either break down when scan to high voltage or show stable high-resistant behavior of traditional amorphous organic films.

To simplify the comparison, the results of the devices studied in this work are summarized in Table 1. It was clearly indicated that the sandwich or multilayer device structure is crucial to the switching characteristics. The switch threshold-applied voltage $V_T$ and on/off ratio are dependent on the TPD/4AAPD thickness ratio to some extent. The lowest threshold voltage and highest on/off ratio were observed in devices with TPD/4AAPD thickness ratio of 1:1. For example, the Ag/TPD (60nm)/4AAPD (120nm)/TPD (60nm)/Ag device has the lowest threshold voltage of ~ 0.8 V and highest on/off ration of ~ $10^4$. Furthermore, we observed that when the TPD layer is thicker than 80 nm, there exists a striking increase in $V_T$ (as high as 8 V) and decrease in the on/off ratio (as low as ~ 2). More importantly, the total organic layer thickness in these devices is similar to that of the sandwich devices with the lowest $V_T$ and highest on/ff ratio. Therefore, the electrical switching performance depends not only on the TPD/4AAPD total layer thickness ratio, but also on each layer thickness. This observation would be very helpful in understanding the switching mechanism as discussed later.

**Table 1  Structure dependence of the electrical switching observed in TPD and 4AAPD devices with Ag electrodes.**

| Active media structure | Layer thickness (nm) | Switch threshold $V_T$ (V) | On/off ratio at 0.2 V |
|---|---|---|---|
| TPD | 50 to 350 | No | No |
| 4AAPD | 50 to 350 | | |
| TPD/4AAPD | 50/50 | | |
| 4AAPD/TPD/4AAPD | 40/20/40 | | |
| TPD/4AAPD/TPD | 15/20/15 | 0.6 ~ 1.2 | $10^2$ ~ $10^4$ |
| | 20/10/20 | 1.0 ~ 1.5 | |
| | 30/30/30 | 0.5 ~ 1.4 | |
| | 60/120/60 | 0.3 ~ 0.8 | |
| TPD/4AAPD/TPD/4AAPD/TPD | 10/10/10/10/10 | 0.6 ~ 1.8 | $10^0$ ~ $10^2$ |
| | 80/50/5/50/15 | 1.5 ~ 6.5 | |
| | 40/60/20/60/80 | 4.5 ~ 8 | |

Similar electrical switching could be easily observed in resemble 4AAPD and TPD sandwich or multilayer devices by using either Au or Cu pair electrodes, indicating that the switching performance is independent of the electrode materials. However, when we use other small dendritic molecules without redox core-shell structures to substitute the embedded 4AAPD layer, no resemble switching behavior could be observed in either sandwich or multilayer devices, suggesting that the redox-gradient and core-shell structures of the 4AAPD molecules are critical to the switches.

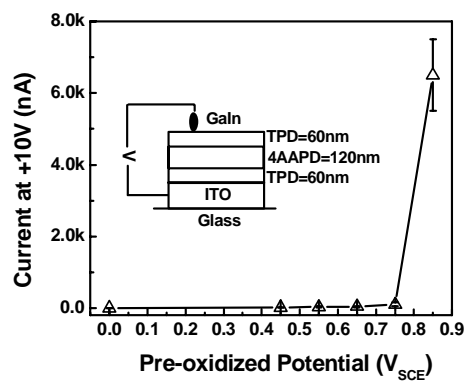

**Fig. 4  Threshold oxidizing potential is observed in TPD and 4AAPD sandwich devices in electrochemical assistant measurements. Inset shows the device structure.**

To understand the switching mechanism, electrochemical assistant pre-oxidation experiment was carried out. As shown in Fig. 4, TPD and 4AAPD sandwich films were deposited onto clean ITO (indium-tin oxide) glass, which was then oxidized in electrolyte solution at different applied potentials by using electrochemical method.[16] After the sample was cleaned and dried, the correlation between the film conductance and the applied potentials was studied by comparing the device current at +10 V. It was shown that there also exists a threshold oxidizing potential, which is about 0.8 $V_{SCE}$. The sample conductance was increased more than 3 orders of magnitude after oxidizing by a threshold potential. Coloration behavior was observed in such oxidized samples, which changed from colorless to deep blue after oxidation. Moreover, the coloration was stable within 3 months in ambient conditions. On the contrary, any low potential, below the threshold, did not induce remarkable increase in the sample conductance (also no coloration was observed in such samples). These observations agree very well with that of the sandwich and multilayer devices shown in Fig. 2 & 3. We thus suggest that the molecular oxidation is responsible for the conductance transition and coloration properties.

All our experimental results suggest that the electrical switching should be resulted from the property changes of the active layers induced by the threshold applied voltage, but not the metal filament formation induced by the top electrode deposition. Furthermore, it was also shown that both the redox-gradient core-shell structure of the embedded 4AAPD dendrimer layer and the special energy well sandwich device structure play a key role in the switching characteristics.

Many papers have been reported on the observation of electrical switching behavior in various systems such as small π-conjugated organic molecules,[17,18,19] organic-metal complexes,[20] and inorganic materials.[21] In these reports, the inner/intra-molecular energy transfers and dynamic dope/dedoping processes are thought to be responsible for the switching effects. In our case, the devices are all-organic, there are neither electron donor/acceptor groups nor embedded metal elements. So the switching mechanism may be different to those reports. In fact, some unusual optical or magnetic properties have been observed in similar sandwich structures, in which the formation of a potential energy well is generally thought to be a critical factor.[22,23] With inspiration from these reports, we think it might be reasonable to explain the observed switching phenomena by a schematic oxidation and charge trapping process. We know that the band gap $E_g$ of 4AAPD is less than that of the TPD.[24] Thus there exists an energy well in the middle 4AAPD layer sandwiched between TPD thin films, where charges could be easily trapped.[1,25] When a external threshold-applied voltage was applied to, efficient core oxidation (and consequent insulator-conductor transitions) of the embedded redox-gradient 4AAPD molecules may be triggered and trapped, respectively. Consequently, electrical switching was observed (see Fig. 5). This simple model can not only explain the device structure dependence of the electrical switching, but also consists with the fact that 4AAPD dendrimers comprised of an electrochemical easy oxidized-core and hard-oxidized shell. According to this model, when the device is constructed with sandwich structures with only one dendrimer layer, single switching should be observed under a threshold-applied voltage. On the contrary, in multilayer devices with several dendrimer layers, multiple step switching should be expected. These explanations are well consistent with the experimental observations.

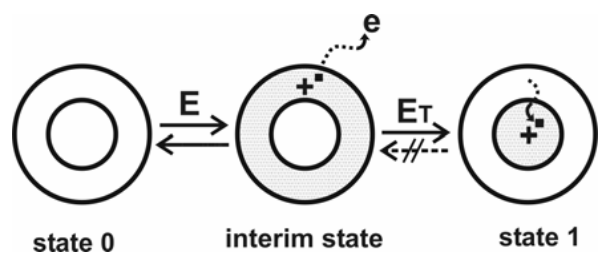

Fig. 5  Schematic core-shell oxidation and charge trapping process of the embedded redox-gradient 4AAPD dendrimer molecule (E – applied electric field and $E_T$ – threshold-applied electric field).

It should be pointed out that we only demonstrate the possibility of using solid-state dendrimer thin films as potential information storage media in this preliminary works. By further extending this strategy, it should be able to find more compatible organic barrier molecules and effective dendrimers so as to lower the threshold voltage and improve the switching performance for practical device applications.[26] Nanoscale charge/information storage could be probably realized on analogous multi-monolayer devices by using conducting atomic force microscopy or scanning tunneling microscopy tip write-read techniques.[27] Detailed studies on such issues are still underway.

In summary, reproducible electrical switching has been observed in dendrimer sandwich devices with active layer constructed with a thin redox-gradient dendrimer layer embedded in TPD thin film when a threshold voltage was applied to. After switching, the device conductivity could be increased more than four orders of magnitude, which can keep stable for more than 50 hrs in ambient conditions. Control experimental results indicated that both the device special energy well structure and the unique redox-gradient of the dendrimers are responsible for the switching observations.

We thank The National Science Foundation for supporting this work through the Materials and Research Science and Engineering Center (grants #DMR-98-09423 and #DMR-02-13985).

[1] Y. Gofer, H. Sarker, J. G. Killian, T. O. Poehler, and P. C. Searson, Appl. Phys. Lett. **71**, 1582 (1997).

[2] K. Sakai, H. Matsuda, H. Kawada, K. Eguchi, and T. Nakagiri, Appl. Phys. Lett. **53**, 1274 (1988).

[3] M. E. Itkis, X. Chi, A. W. Cordes, and R. C. Haddon, Science **296**, 1443 (2002).

[4] J. C. Li, Z. Q. Xue, X. L. Li, W. M. Liu, S. M. Hou, Y. L. Song, L. Jiang, D. B. Zhu, X. X. Bao, and Z. F. Liu, Appl. Phys. Lett. **76**, 2532 (2000).

[5] C. K. Song, B. W. Koo, and C. K. Kim, Jpn. J. Appl. Phys. **41**, 2735 (2002).

[6] J. P. J. Markham, S. C. Lo, S. W. Magennis, P. L. Burn, and I. D. W. Samuel, Appl. Phys. Lett. **80**, 2645 (2002).

[7] K. Yamamoto, M. Higuchi, S. Shiki, M. Tsuruta, and H. Chiba, Nature (London) **415**, 509 (2002).

[8] A. Archut, G. C. Azzellini, V. Balzani, L. D. Cola, and F. Vogtle, J. Am. Chem. Soc. **120**, 12187 (1998).

[9] D. M. Junge and D. V. McGrath, Chem. Commun. **9**, 857 (1997).

[10] J. W. Weener and E. W. Meijer, Adv. Mater. **12**, 741 (2000).

[11] M. W. P. L. Baars, M.C.W. van Boxtel, C.W.M. Bastiaansen, D.J. Broer, S.H.M. Sontjens, and E.W. Meijer, Adv. Mater. **12**, 715 (2000).

[12] N. Tessler, V. Medvedev, M. Kazes, S.H. Kan, and U. Banin, Science **295**, 1506 (2002).

[13] T. D. Selby, K.-Y. Kim, and S. C. Blackstock, Chem. Mater. **14**, 1685 (2002).

[14] K.-Y. Kim, J. D. Hassenzahl, T. D. Selby, G. J. Szulczewski, and S. C. Blackstock, Chem. Mater. **14**, 1691 (2002).

[15] L. L. Miller, R. G. Duan, D. C. Tully, and D. A. Tomalia, J. Am. Chem. Soc. **119**, 1005 (1997).

[16] Experimental details on the electrochemical oxidation process have been reported in our previous works. Here, ITO glass was used as the bottom electrode so as to detect the film color changes before and after oxidation. GaIn liquid metal microscale tips were used as top electrode in order to prevent metal filament formation from top electrode deposition.

[17] J. M. Seminario, P. A. Derosa, and J. L. Bastos, J. Am. Chem. Soc. **124**, 10266 (2002).

[18] Chao Li, Daihua Zhang, Xiaolei Liu, Song Han, Tao Tang, Chongwu Zhou, Wendy Fan, Jessica Koehne, Jie Han, Meyya Meyyappan, A. M. Rawlett, D. W. Price, and J. M. Tour, Appl. Phys. Lett. **82**, 645 (2003).

[19] A. Rochefort, R. Martel, and P. Avouris, Nano Letters **2**, 877 (2002).

[20] L. P. Ma, J. Liu, and Y. Yang, Appl. Phys. Lett. **80**, 2997 (2002).

[21] D. C. S. Souza, V. Pralong, A. J. Jacobson, and L. F. Nazar, Science **296**, 2012 (2002).

[22] H. B. Nie, C. K. Ong, J. P. Wang, and Z. W. Li, J. Appl. Phys. **93**, 7252 (2003).

[23] Z. Y. Xie, J. S. Huang, C. N. Li, S. Y. Liu, Y. Wang, Y. Q. Li, and J. C. Shen, Appl. Phys. Lett. **74**, 641 (1999).

[24] J. C. Li, K.-Y. Kim, S. C. Blackstock, and G. J. Szulczewski, submitted Chem. Mater.

[25] C. -Y. Liu, H. L. Pan, M. A. Fox, and A. J. Bard, Chem. Mater. **9**, 1422 (1997).

[26] M. R. Bryce, P. D. Miguel, and W. Devonport, Chem. Commun. **23**, 2565 (1998).

[27] J.C. Li, Z. Q. Xue, W. M. Liu, S. M. Hou, X. L. Li, and X. Y. Zhao, Phys. Lett. A **266**, 441 (2000).